\documentclass[11pt,a4paper]{article}

\usepackage{graphicx}
\usepackage{booktabs}
\usepackage{multirow}
\usepackage{microtype}
\usepackage{authblk}
\usepackage{url}
\usepackage{listings}
\usepackage{color}

\definecolor{dkgreen}{rgb}{0,0.6,0}
\definecolor{gray}{rgb}{0.5,0.5,0.5}
\definecolor{mauve}{rgb}{0.58,0,0.82}

\begin{document}

\title{JavaNPST: Nonparametric Statistical Tests in Java}
\author[1]{Joaqu\'{\i}n Derrac\thanks{jderrac@decsai.ugr.es}}
\author[2]{Salvador Garc\'{\i}a\thanks{salvagl@decsai.ugr.es}}
\author[2]{Francisco Herrera\thanks{herrera@decsai.ugr.es}}
\affil[1]{Affectv: Affectv Limited, 33-34 Alfred Place, London, WC1E 7DP, United Kingdom.}
\affil[2]{Department of Computer Science and Artificial Intelligence, University of Granada, 18071, Granada, Spain.}

\providecommand{\keywords}[1]{\textbf{\textit{Keywords---}} #1}

  \maketitle

\begin{abstract}
Nonparametric statistical tests are useful procedures that can be applied in a wide range of
situations, such as testing randomness or goodness of fit, one-sample, two-sample and multiple-sample analysis, association between bivariate samples or count data analysis. Their use is often preferred to parametric tests due to the fact that they require less restrictive assumptions about the population sampled.

In this work, JavaNPST, an open source Java library implementing 40 nonparametric statistical tests, is presented. It can be helpful for programmers and practitioners interested in performing nonparametric statistical analyses, providing a quick and easy way of running these tests directly within any Java code. Some examples of use are also shown, highlighting some of the more remarkable capabilities of the library.
\end{abstract}


\lstset{language=Java}

\section{Introduction}

Nonparametric statistical tests \cite{Higgins03,Sheskin11} comprise a class of hypothesis testing procedures in which the null hypothesis is not a statement about parameter values. Instead, the hypotheses are usually concerned with the probability distribution of the sample data used in the test or with the form of the population. Generally speaking, most of them can be considered to be \textit{distribution-free} procedures, in the sense that the distribution of the random variables involved does not depend on the specific distribution function of the population from which the testing sample was drawn \cite{Gibbons10}.

There is a notable number of problems that can be tackled with these procedures. Inside the nonparametric statistical inference field, a practitioner can find a relatively high number of tasks such as testing for randomness or goodness of fit, locating the median or some other quantile of a particular distribution, comparing  two samples in terms of location of medians or scale, testing equality of multiple independent samples or analyzing count data.

Interest in this topic is widespread: For example, a quick search in \url{www.amazon.com} with the search query \textit{non-parametric statistical tests} would return more than 800 different books. The same search, performed in Google Scholar would retrieve more than 268,000 different documents indexed on the Web. From statisticians developing new procedures and analyzing their features, to practitioners in a very broad range of fields, nonparametric statistical tests have attracted the attention of the research community since the beginning of the 20th century.

Currently, many statistical software suites such as SPSS \cite{SPSS}, SAS \cite{SAS}, Minitab \cite{Minitab} or StatXact \cite{StatXact, Mehta91} include some of these procedures. Many minor software packages can be found on the Web implementing some of the most popular methods in various programing languages like C, C++, Java, Fortran90, R, Matlab or Mathematica. However, to the best of our knowledge, there is not a single software package implementing a complete set of nonparametric statistical tests ready to be used in most situations. Often, practitioners willing to employ nonparametric tests in their applications need to search through very different sources until they get the specific set of tests needed, or to rely on the tests implemented in other software suites, outside of their own software projects.

In this paper we present JavaNPST, a Java library of nonparametric statistical tests, which is suitable for practitioners in most of these situations. It is an open source library, featuring 40 different nonparametric tests (freely available at \url{http://sci2s.ugr.es/software/javanpst/}). They are classified in 10 families of methods, each one oriented to tackle a particular kind of problem. As it is a Java library, it can be easily integrated to any Java software project without requiring a deep understanding of the language, and can be used under any operating system able to run the Java Virtual Machine. Moreover, given the popularity of Java, many solutions for translating/running Java code in other environments have been developed, such as rJava in R \cite{Rlanguage}, which increases the accessibility of our library to practitioners interested in its use. Therefore, it should not be difficult to integrate it into existing software projects, even if developed in a different programming language.

The design of the library offers a simple interface to the user. A homogeneous set of methods, common to every procedure, allows the user to define the necessary tests, set them up and run them, and to obtain complete results including test statistics and computed $p$~values. This property, together with the generality of problems covered by the 10 families, makes JavaNPST a suitable tool in many different application fields, even to be used by teachers in introductory courses about nonparametric statistical inference, as a resource to let students experiment with the tests by themselves.

The rest of this paper is organized as follows: Section~\ref{sec:theory} surveys the 10 families of nonparametric tests considered in the library. Section~\ref{sec:library} describes the structure and the main features of JavaNPST. Section~\ref{sec:examples} shows several cases of use in various application fields. Finally, Section~\ref{sec:discussion} discusses the conclusions achieved.

\section{Nonparametric statistical tests}\label{sec:theory}

Given their broad definition, a substantial number of tests have been developed for many situations in which nonparametric statistical inference fits the problem that has arisen (mostly due to the nature of the data). Thus, a proper way of giving a quick snapshot of the field would be to establish a taxonomy of the tests, classified by the kinds of problems that they tackle. In the development of JavaNPST, we have followed the taxonomy established in \cite{Gibbons10}, where 10 families of methods are presented (Table~\ref{tab:families} summarizes them):

\begin{table}[!ht]
	\centering
	\resizebox{0.85\textwidth}{!}{
	\begin{tabular}{clc}\toprule
	Family & Test & Reference \\ \midrule
	\multirow{4}{*}{Tests of randomness} & Number of Runs & \cite{Swed43} \\
	& Runs Up and Down & \cite{Edgington61}\\
	& Runs Up and Down (Median) & \cite{Swed43} \\
	& Von Neumann & \cite{Bartels82}\\ \midrule
	\multirow{4}{*}{Tests of goodness of fit} & Chi-Square test & \cite{Pearson00} \\
	& Kolmogorov-Smirnov & \cite{Smirnov39} \\
	& Lilliefors & \cite{Lilliefors67} \\
	& Anderson-Darling & \cite{Anderson52} \\ \midrule
	& Confidence Quantile & \cite{Gibbons10} \\	
	One-sample and & Population Quantile & \cite{Gibbons10} \\	
	paired-samples & Sign test & \cite{Sheskin11} \\	
	& Wilcoxon Signed-Ranks & \cite{Wilcoxon45} \\ \midrule	
	& Wald-Wolfowitz & \cite{Wald40} \\	
 	 Two-Sample & Median test & \cite{Brown48} \\	
 	 general procedures & Control Median & \cite{Mathisen43} \\	
 	& Kolmogorov-Smirnov & \cite{Smirnov39} \\ \midrule	
	\multirow{2}{*}{Location problem} & Wilcoxon Rank-Sum & \cite{Wilcoxon45} \\	
 	& van der Waerden& \cite{Waerden52} \\	\midrule	
 	\multirow{6}{*}{Scale problem} & David-Barton & \cite{David58} \\	
 	& Freund-Ansari-Bradley & \cite{Ansari60} \\	
 	& Mood & \cite{Mood54} \\	
 	& Klotz & \cite{Klotz62} \\	
 	& Siegel-Tukey & \cite{Siegel60} \\	
 	& Sukhatme & \cite{Sukhatme57} \\	\midrule	
 	& Extended Median test & \cite{Brown48} \\	
 	Equality of & Kruskal-Wallis & \cite{Kruskal52} \\	
 	independent samples & Jonckheere-Terpstra & \cite{Jonckheere54} \\	
 	& Charkraborti-Desu	& \cite{Chakraborti88} \\ \midrule
 	Association for & Kendall & \cite{Kendall70} \\
 	bivariate samples & Daniel Trend & \cite{Kendall70} \\ \midrule	
 	& Friedman & \cite{Friedman37} \\
 	Association in & Page & \cite{Page63} \\
 	\multirow{2}{*}{multiple classifications} & Concordance Coefficient & \cite{Gibbons10} \\
 	& Incomplete Concordance & \cite{Durbin51} \\
 	& Partial Correlation & \cite{Maghsoodloo75} \\ \midrule	
	& Contingency Coefficient & \cite{Pearson04} \\
 	Analysis of & Fisher's exact test & \cite{Fisher22} \\
 	\multirow{2}{*}{count data} & McNemar & \cite{McNemar47} \\
 	& Multinomial Equality test & \cite{Gibbons10} \\
 	& Ordered Equality test & \cite{Gibbons10} \\
	 \bottomrule
	\end{tabular}
    }
	\caption{Tests included in the current version of JavaNPST.}
	\label{tab:families}
\end{table}

\begin{itemize}

	\item \textbf{Tests of randomness}: These tests are used to check randomness either in binary symbolic sequences (for example XYXYXY) or in numerical sequences. The Number of Runs test can be applied to test randomness of the former, whereas, for numerical sequences, a Runs Up and Down test based either on previous values or on the median value of the sequence can be used. Furthermore, the Von Neumann ranks-based test can also be applied to numerical sequences.
	
	\item \textbf{Tests of goodness of fit}: JavaNPST implements the Chi-Square test for the adjustment of data to discrete distributions. For continuous fit, the Kolmogorov-Smirnov test is provided as an omnibus test that should be useful in most situations. Two more tests, Lilliefors and Anderson-Darling, can also be applied when the adjustment has to be tested against a normal or exponential distribution with unknown parameters.	
		
	\item \textbf{One-sample and paired-sample procedures}: The usefulness of these procedures lies in verifying hypotheses related to a given quantile of the sample's distribution (usually the median). The Confidence Quantile test is used to obtain a confidence interval for a specified quantile, whereas the Population Quantile test allows the user to test a hypothesis concerning a specific value for any quantile. For paired samples, the Sign test can be employed to test the location of the median of the population of differences. Finally, the Wilcoxon Signed-Ranks test is used in the same scenario, but employing more information concerning relative magnitudes as well as directions or differences.
	
	\item \textbf{Two-sample general procedures}: These procedures are used to verify equality between two samples, without assuming any specific model. The Wald-Wolfowitz test maps the values of the ordered combined sample into a binary symbolic sequence, and then applies a Number of Runs test to detect differences if too few runs are found. Median and Control Median tests highlight differences between populations using the median value of the samples. Finally, the Kolmogorov-Smirnov Two-Sample test can also be used to test equality, under general assumptions.

	\item \textbf{Tests for the location problem}: Tests in this category follow the location model
	
	\begin{eqnarray}
		H_{0}:F_{Y}(x) = & F_{X}(x) & \textrm{ for all } x \\ \nonumber
		H_{L}:F_{Y}(x) = & F_{X}(x-\theta) & \textrm{ for all } x \textrm{ and some } \theta \neq 0
		\label{eq:locationModel}
	\end{eqnarray}
	
	According to this model, the Wilcoxon Rank-Sum tests provide a way to compare two samples and estimate a confidence interval for the location parameter. The van der Waerden test also follows the location model, but using inverse normal scores as weights when forming the linear ranks statistic of the test.
		
	\item \textbf{Tests for the scale problem}: Analogous to the former category, the tests for the scale problem check differences between both distributions regarding the scale parameter
	
	\begin{eqnarray}
		H_{0}:F_{Y}(x) = & F_{X}(x) & \textrm{ for all } x \\ \nonumber
		H_{S}:F_{Y}(x) = & F_{X}(\theta x) & \textrm{ for all } x \textrm{ and some } \theta > 0 , \theta \neq 1
		\label{eq:scaleModel}
	\end{eqnarray}
	
	Following this model, the David-Barton, Freund-Ansari-Bradley and Mood tests establish three different schemes of weights to obtain linear ranks for testing the hypothesis. On the other hand, the Klotz test obtains a set of ranks based on the van der Warden test, whereas the Siegel-Tukey test employs the Wilcoxon Rank-Sum test weights and the Sukhatme procedure tests the hypothesis with a Mann-Whitney based statistic.
	
	\item \textbf{Tests of equality of independent samples}: This family contains several methods oriented to tackling the natural extension of the Two-Sample problem, namely the $k$-sample problem, whose null hypothesis states that all $k$ samples are drawn from identical populations, whereas the general alternative is simply that the populations differ in some way
	
	\begin{eqnarray}
		H_{0}: & F_{1}(x) = F_{2}(x) = \cdots = F_{k}(x) & \textrm{ for all } x \\ \nonumber
		H_{1}: & F_{i}(x) \neq F_{j}(x) & \textrm{ for at least one } i \neq j
		\label{eq:equalityModel}
	\end{eqnarray}
	
	The Extended Median test and the Kruskal-Wallis test are in this category, as the natural extensions of the Median and the Wilcoxon Rank-Sum tests, respectively.
	
	Ordered alternatives concerning the $\theta$ location parameter, such as
	
	\begin{equation}
		H_{1}: \theta_{1} \leq \theta_{2} \leq \cdots \leq \theta_{k}
		\label{eq:orderedAlternatives}
	\end{equation}
	
	can be tested with the Jonckheere-Terpstra test. Finally, comparisons with a control (where the null hypothesis states that every $\theta$ parameter is equal to or greater than that of the control, with at least one of the inequalities strict) can be performed with the Charkraborti-Desu test, the last procedure of this family.	
	
	\item \textbf{Measures of association for bivariate samples}: Two well-known measures of association for bivariate samples, together with their respective formulation of hypothesis for association, are included in this category: The Kendall's $\tau$ coefficient (Kendall's test) and Spearman's correlation coefficient (Daniel Trend test).
	
	\item \textbf{Measures of association in multiple classifications}: These procedures are the nonparametric analogs of the two-way analisys-of-variance problem, where data cannot be considered as single random sample because of certain relationships between them, such as columns and row effects. The general test to employ here is the Friedman test, whereas the Page test is used for testing ordered alternatives (with the same form as the alternative shown in Equation~\ref{eq:orderedAlternatives}).
	
	Other tests in this category can be used to find a measure of strength of the relationship between rankings. This measure, the coefficient of concordance, can be computed for complete samples (the Concordance Coefficient procedure), or for incomplete samples (the Incomplete Concordance procedure) belonging to Youden squares or Latin squares experiment designs. Finally, partial correlation between ranks (Partial Correlation test) can also be computed using Kendall's $\tau$.
	
	\item \textbf{Analysis of count data}: Count data can also be analyzed in various ways using nonparametric procedures. Several coefficients of no association between rows and columns can be computed for $n$ $x$ $m$ contingency tables (Contingency Coefficient procedure).
	
	Fisher's exact test consists of testing the significance of the association between classifications in 2 $x$ 2 contingency tables. McNemar's test for 2 $x$ 2 contingency tables can be applied to determine whether the row and column marginal frequencies are equal.
	
	Finally, the multinomial test of equality is used to test the equality of probabilities in multiple classifications. The alternative hypothesis may simply be that the probabilities are not equal (the Multinomial Equality test) or that they form an ordered alternative (the Ordered Equality test).

\end{itemize}

\section{The JavaNPST library}\label{sec:library}

JavaNPST is a Java library featuring a wide collection of nonparametric tests, together with several definitions of data structures and numeric distributions needed to deploy and carry out the tests. Public classes belonging to the library feature interfaces are restricted to the essentials, thus facilitating its use inside high-level applications.

The library is built around three core packages:

\begin{itemize}

	\item Data (javanpst.data): Modeling the data structures of the library (sequences and tables).
	
	\item Distributions (javanpst.distributions): Including classical discrete and continuous distributions, and distributions related to the tests.
	
	\item Tests (javanpst.tests): Implementing the nonparametric tests of the library.

\end{itemize}

A fourth package, Utils (javanpst.utils) contains some common tools for the internal use of the library (functions for input/output of files, formats, some mathematical operations, and so forth).

All these elements have been developed using an object-oriented style. Therefore, the user of JavaNPST can expect to find objects representing all the necessary pieces to perform an analysis. In this way, a typical use of the library will include the declaration of an object modeling the data to analyze (data sequences, or samples from various populations), the creation of another object representing the nonparametric test selected, and, finally, the evaluation of the test, obtaining an output report as a result.

The rest of this section is devoted to describing in depth the main packages: The Data package (Section~\ref{sub:data}), the Distributions package (Section~\ref{sub:distributions}) and the Tests package (Section~\ref{sub:tests}). A more detailed description of the JavaNPST API, together with other resources and examples of use can be found at the web page of the project \url{http://sci2s.ugr.es/software/javanpst/}.

\subsection{The Data package}\label{sub:data}

Before running a test, data should be provided to the test object. Input data - usually composed of various samples drawn from several populations, or of sequences of numerical or string values - should be represented in a proper way, before starting its analysis.

JavaNPST defines two main data structures for storing samples: Sequence and DataTable objects. The former is suitable when the sample to study is composed either of a sequence of values or of a sample drawn from a single population. The latter should be used when data represents more than one sample.

\subsubsection{Sequences}

Sequences can instantiated as two different objects: NumericSequence (for numerical data) and StringSequence (for textual data). Both can be easily built from ArrayList objects:

\begin{lstlisting}
// Let 'array1' be an ArrayList<double> already filled,
// and 'array2' an ArrayList<String>, also filled

NumericSequence sequence1 = new NumericSequence(array1);
StringSequence sequence2 = new StringSequence(array2);
\end{lstlisting}

A second option for initializing Sequence objects is to load data into them directly from a file. Currently, JavaNPST allows reading data to be stored in XML, CSV and TXT formats. For both kinds of sequences, the input format is the following \footnote{Elements do not have to be separated by new lines}:

\begin{itemize}
	\item XML:
	\begin{lstlisting}
	<sequence>
		<element>Value 1</element> ...
		<element>Value 2</element> ...
		...
		<element>Value K</element> ...
		...
		<element>Value N</element> ...
	</sequence>
	\end{lstlisting}
	
	\item CSV or TXT:
	\begin{lstlisting}
	Value 1;Value 2;...;Value K;
	... Value N
	\end{lstlisting}		
\end{itemize}

Once the data file has been properly formatted, new sequences can be created as follows

\begin{lstlisting}
NumericSequence sequence1 = new NumericSequence();
StringSequence sequence2 = new StringSequence();

sequence1.readXML(pathToXMLFile);
sequence1.readCSV(pathToCSVFile);
\end{lstlisting}

\subsubsection{Data tables}

The DataTable object can be used to store numerical data in a tabular way. Usually, data tables in JavaNPST are used to store several samples drawn from different populations, where values in a column belong to the same sample \footnote{Unless explicitly stated in the documentation of a test}.

In a similar way to sequences, data tables can be filled easily during their instantiation

\begin{lstlisting}

double matrix [][] = {{1,2,2.5},
                      {1.2,3,2},
                      {1.5,2.7,4.9},
                      {1,2.2,2.1}};
											
DataTable table = new DataTable(matrix);
\end{lstlisting}

Data tables can also load data from XML, CSV or TXT files. The required format is shown below

\begin{itemize}

	\item XML:
	
	\begin{lstlisting}
	<tabular rows = ''#Rows'' columns = ''#Columns''>
		<row><element>Value(1,1)</element>...</row>
		...
		<row><element>Value(n,1)</element>...</row>
	</tabular>
	\end{lstlisting}
		
	\item CSV or TXT:
	
	\begin{lstlisting}
	Value(1,1);...;Value(1,n)
	...
	Value(n,1);...;Value(m,n)
	\end{lstlisting}
	
\end{itemize}

Again, loading data from a file into a DataTable object can be done in a single instruction

\begin{lstlisting}
DataTable table = new DataTable();

table.readXML(pathToXMLFile);
\end{lstlisting}

Apart from builders and methods for reading data, Sequence and DataTable objects have an interface providing basic functionality for obtaining and modifying values, writing contents to a file and so forth, available to the user \footnote{For more details, see the JavaDoc documentation of JavaNPST}. Although data structures are automatically managed by the test objects - and hence, user intervention is not needed - data manipulation functions provided by the interface of these classes may be useful to assess data if, for example, multiple tests are going to be carried out over the same samples.

The rest of the Data package contains several inner classes developed for writing and loading data, as well as specific data structures used for storing data belonging to test distributions. 

\subsection{The Distributions package}\label{sub:distributions}

This package contains implementations of all the distributions used by the tests of the library. They can be classified as

\begin{itemize}

	\item Common distributions
	
	\begin{itemize}

		\item Discrete distributions: BinomialDistribution, PoissonDistribution, $\ldots$ 	

		\item Continuous distributions: NormalDistribution, ChiSquareDistribution, $\ldots$ 	
		
	\end{itemize}
	
	\item Tests distributions: KolmogorovDistribution, KendallDistribution, $\ldots$
	
\end{itemize}

Usually, a programmer using JavaNPST should not need to declare or employ a distribution in isolation, since they are automatically created and configured as soon as they are required by each test. However, direct access is provided as sometimes it may be useful to use some common distributions directly. An example, illustrating how a normal distribution $N\left(5,0.3\right)$ can be modeled, is shown as follows

\begin{lstlisting}
NormalDistribution normal = new NormalDistribution();

normal.setMean(5);
normal.setSigma(0.3);
\end{lstlisting}

With the distribution initialized, some operations such as, for example, the computation of cumulative distribution values, are straightforward

\begin{lstlisting}
double p;
double Z = 1.35;

p = normal.computeCumulativeProbability(Z);
\end{lstlisting}

On the other hand, test related distributions are used for internal management by those tests whose exact distribution is known but is stored inside the library as a table (Wilcoxon's or Lilliefors' distribution are representative examples). As they are managed internally, users should not need to access them in most cases.

\subsection{The Tests package}\label{sub:tests}

Test objects are the main components on which JavaNPST is based. They allow the user to, given an appropriately formatted set of data, perform a test and get all the results obtained in the inference process.

All the tests share a common interface which provides the user with a basic functionality. In this way, setData and doTest methods can be used to load the data to test and perform the inference, respectively, whereas the printReport method shows the full results of the process.

As an example, when data have already been represented in a DataTable, a Median test can be performed using only these methods

\begin{lstlisting}
DataTable data = new DataTable(samples);
MedianTest median = new MedianTest();

median.setData(data);
median.doTest();

System.out.println(median.printReport());
\end{lstlisting}

Moreover, in some scenarios the user could be interested in obtaining a single $p$~value of the inference (that is, only the $p$~value of a tail), or the value of a test statistic. JavaNPST also allows this possibility, offering direct access to every output value of each test

\begin{lstlisting}
double leftTailExactPValue;

DataTable data = new DataTable(samples);
MedianTest median = new MedianTest();

median.setData(data);
median.doTest();

leftTailExactPValue = median.getExactRightPValue();
\end{lstlisting}

No more procedures or operations are needed to perform a test and obtain its results. This simplicity is one of the key features of JavaNPST, making the task of integrating the tests in the development of more advanced software projects very easy.

\section{Using JavaNPST: Examples of use}\label{sec:examples}

JavaNPST is freely available at \url{http://sci2s.ugr.es/software/javanpst/}. On this website, users can find both the JAR file with the library and the API documentation, in HTML (JavaDoc based) format. Moreover, the source code is also offered, under the terms of the GNU Public License GPL.V3.

The source code itself is documented thoroughly, thus practitioners should not find any problem in using the 40 tests of the library and its associated tools. In addition, the web site also provides a number of code samples illustrating the use of the tests, with various examples of their use (see \url{http://sci2s.ugr.es/software/javanpst/tests.php}).

In this section, we will focus our attention on two elaborate examples. Section~\ref{sub:machineLearning} covers the use of nonparametric tests as a tool for contrasting comparisons of machine learning algorithms (reviewing an already published research experiment), and Section~\ref{sub:credit} shows an application of the Kolmogorov-Smirnov test of JavaNPST for optimizing a scoring model in credit risk management.

\subsection{Nonparametric tests for comparing machine learning algorithms}\label{sub:machineLearning}

In recent years, nonparametric tests have attracted the attention of numerous researchers in the machine community. In this field, nonparametric tests arise as a tool for contrasting results of experiments. In most cases, their use is preferred to parametric alternatives ---such as the $t$-test--- due to the impossibility of fulfilling the necessary conditions for safety (independence, normality and homoscedasticity \cite{Garcia09}).

\begin{table}[t]
	\centering
	\resizebox{\textwidth}{!}{
	\begin{tabular}{lcc@{\hspace{40pt}}lcc} \toprule
Data set & DROP3 & CHC & Data set & DROP3 & CHC \\ \midrule	
abalone & 0.7550 & 0.5517 & lrs & 0.8607 & 0.8211 \\
anneal & 0.8735 & 0.9255 & lymphography & 0.6687 & 0.8321 \\
audiology & 0.7255 & 0.8216 & new-thyroid & 0.8727 & 0.8330 \\
autos & 0.5946 & 0.8297 & optdigits & 0.8997 & 0.5581 \\
balance & 0.7821 & 0.8723 & page-blocks & 0.9562 & 0.7740 \\
breast-cancer & 0.8186 & 0.9178 & pendigits & 0.9468 & 0.5089 \\
cancer & 0.9362 & 0.9451 & phoneme & 0.8191 & 0.7262 \\
card & 0.7468 & 0.9163 & pima & 0.7775 & 0.8708 \\
dermatology & 0.8615 & 0.9021 & post-operative & 0.8790 & 0.7852 \\
ecoli & 0.8380 & 0.8673 & primary-t & 0.7369 & 0.8075 \\
gene & 0.6230 & 0.568 & promoters & 0.6042 & 0.8187 \\
german & 0.7441 & 0.8553 & satimage & 0.8737 & 0.5889 \\
glass & 0.7078 & 0.8471 & segment & 0.8648 & 0.9049 \\
glass-g2 & 0.6783 & 0.8667 & sick & 0.9483 & 0.7957 \\
heart & 0.7889 & 0.9198 & sonar & 0.6670 & 0.8649 \\
heart-c & 0.7937 & 0.8658 & soybean & 0.8132 & 0.9057 \\
hepatitis & 0.8521 & 0.9186 & texture & 0.9015 & 0.5130 \\
horse & 0.7872 & 0.9125 & tic-tac-toe & 0.7316 & 0.8838 \\
hypothyroid & 0.9749 & 0.7834 & vehicle & 0.6808 & 0.8155 \\
ionosphere & 0.9070 & 0.8592 & vote & 0.9072 & 0.9265 \\
iris & 0.8348 & 0.9141 & vowel & 0.5741 & 0.8094 \\
Kr-vs-Kp & 0.7623 & 0.7926 & waveform & 0.7291 & 0.5044 \\
labor & 0.6308 & 0.8846 & wine & 0.8298 & 0.8801 \\
led24 & 0.5583 & 0.7894 & yeast & 0.7282 & 0.8374 \\
liver & 0.5871 & 0.9061 & zoo & 0.7681 & 0.8374 \\ \bottomrule
	\end{tabular}
}
	\caption{Performance results in (1-Storage requirements) measure of DROP3 and CHC algorithms. Adapted from Table 2 (pp 397-398) in \cite{GarciaPedrajas10}.}
	\label{tab:machine}
\end{table}

A representative example can be found in \cite{GarciaPedrajas10}, where the Sign and Wilcoxon tests are used to contrast the results obtained when analyzing the behavior of several machine learning algorithms. Here we reproduce one of the comparisons performed, between the DROP3 and CHC methods, in terms of a specific performance measure (storage requirements).

Table~\ref{tab:machine} shows an adaptation of the values used for the comparison (values have been changed to ($1.0-$storage requirements), to aim for maximization of the performance measure before applying Wilcoxon and Sign tests).

Using JavaNPST, Wilcoxon and Sign tests are carried out in a few lines:

\begin{lstlisting}
//Data are stored in a XML file
String file = "./drop3VSchc.xml";
		
//Create data structure
DataTable data = new DataTable();
		
data.readXML(file);

//Create tests
SignTest sign = new SignTest(data);
WilcoxonTest wilcoxon = new WilcoxonTest(data);
		
//Run procedure
sign.doTest();
		
//Print two-tailed exact p-value
System.out.println("Sign test (double tail exact p-value): "
	+ String.format("%.4f%n",sign.getExactDoublePValue()));
		
//Run procedure
wilcoxon.doTest();
		
//Print results
System.out.println("Wilcoxon test (double tail exact p-value): "
	+ String.format("%.4f%n",wilcoxon.getExactDoublePValue()));
\end{lstlisting}
\begin{lstlisting}
==============================================================

Sign test (double tail exact p-value): 0.0066
Wilcoxon test (double tail exact p-value): 0.0724
\end{lstlisting}

The results obtained exactly match those reported for the Sign test by the authors ($p$~value: 0.0066) and shows a very near $p$~value for the Wilcoxon test ($p$~value: 0.0718). With these results we have corroborated the study performed (with small differences probably caused by their use of an asymptotic distribution of the Wilcoxon statistic instead of the exact one) and shown how JavaNPST can be used to perform these tests in a quick and easy way.

\subsection{Nonparametric tests in credit scoring improvement}\label{sub:credit}

It is also possible to find applications of nonparametric tests in the field of Economics. Specifically, in this section we shall show how the Kolmogorov-Smirnov Two-Sample test (and its associated statistic) can be used to evaluate the quality of a scoring process, and how JavaNPST can be employed to develop a testing method for them.

Credit scoring \cite{Siddiqi05} is the process by which the potential risk posed by lending money to consumers is evaluated, with the aim of mitigating losses due to bad debt. In this way, a great variety of techniques may be used to establish a scoring scale, which will be used to discriminate between ``goods'' credits (those which are very likely to be returned) and ``bads'' ones (those which will be lost). Hence, new petitions can be segmented between those two classes, depending on the score that they achieve.

\begin{figure}[th]
	\centering
		\includegraphics[scale=0.9]{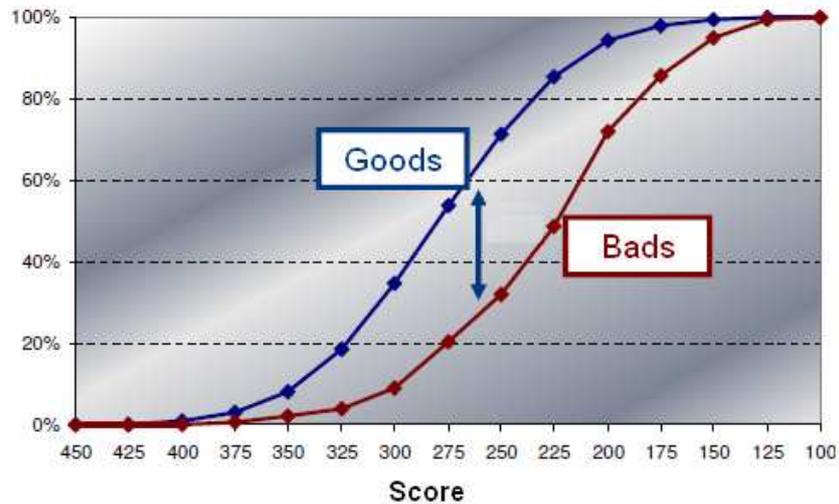}
		\caption{Differences between cumulative distributions of ``goods'' and ``bads'' credits. The Kolmogorov-Smirnov statistic is estimated by measuring the maximum separation between distributions.}
		\label{fig:credit}
\end{figure}

One way of analyzing the predictive power of a credit score system (such as, for example, a credit card) consists of using the Kolmogorov-Smirnov statistic to measure how far apart the cumulative distribution functions of the scores of the ``goods'' and the ``bads'' credits are. If the Kolmogorov-Smirnov statistic is high enough, both distributions will be well discriminated, and thus the predictive power of the scoring system will be good. A graphical example is shown in Figure~\ref{fig:credit}, where the vertical line marks the maximum separation between both distributions, representing the Kolmogorov-Smirnov statistic.

The process of optimizing such a scoring model can be evaluated using JavaNPST. Below we show an example where, given a SearchMethod able to generate several ScoreCard models (inside an optimization process), classifications obtained by these models are evaluated by a Kolmogorov-Smirnov Two-Sample test (K\_STest), using the Kolmogorov-Smirnov statistic as a performance measure. When all the iterations of the SearchMethod have been carried out, the best model is reported, as well as its associated statistic and the $p$~value of the difference between the distributions of ``goods'' and ``bads'' credits

\begin{lstlisting}
ScoreCard model, bestModel;
SearchMethod method;
double modelClassification [][];	
K_STest test = new K_STest();
DataTable tableScore = new DataTable();
double maximumKS = 0;

while(method.hasIterations()){
	
	model = method.searchNewModel();
	modelClassification = model.classifyCredits();
	test.setData(modelClassification);
	test.doTest();
	
	if(maximumKS < test.getDn()){
		maximumKS = test.getDn();
		bestModel = model
	}
}

modelClassification = bestModel.classifyCredits();
test.setData(modelClassification);
test.doTest();
System.out.println(``Best model found. KS statistic: '' + test.getDn());
System.out.println(``P-Value: '' + test.getExactDoublePValue());
\end{lstlisting}

This is a good example of an improvement of test libraries (such as JavaNPST) over software suites: Partial results of the tests can be easily included inside a greater process, without requiring the development of complex procedures to extract data, migrate it to a software suite, and send the results to the native application.

\section{Conclusions}\label{sec:discussion}

In this article we have introduced JavaNPST, a software library featuring 40 nonparametric tests which can be applied in 10 different families of problems. Given the homogeneity of use of the tests, its easy integration and the wide range of problems covered, it may become a useful tool for practitioners in very different fields of science and research, providing them with an open source library ready to be used in any Java software project where a nonparametric statistical test is necessary.

Finally, we conclude this work noting that, despite the current version of JavaNPST providing tests that fit many situations, the possibility of extending it is open: with the main infrastructure of data structures, distributions, and tests already working, inclusion of new procedures as needed should be straightforward, thus enabling it to remain up-to-date.

\section*{Acknowledgments}
Supported by the Spanish Ministry of Science and Technology. This work was supported by the Research Project TIN2011-28488.

\bibliographystyle{unsrt}
\bibliography{javanpst}

\begin{thebibliography}{10}
\bibitem{Pearson00}
K.~Pearson.
\newblock On the criterion that a given system of deviations from the probable
  in the case of a correlated system of variables is such that it can be
  reasonably supposed to have arisen from random sampling.
\newblock {\em Philosophical Magazine}, 50(5):157--175, 1900.
\end{thebibliography}

\end{document}